\documentclass{egpubl}
\usepackage{pg2025}

\usepackage{amsmath}
\usepackage{algorithm}
\usepackage{algpseudocode}
\usepackage{float}
\usepackage{color}
\usepackage{enumitem}
\usepackage{multirow}
\usepackage{amssymb}
\usepackage{array} % For extended column formatting
\usepackage{amssymb} % For checkmarks
\usepackage{xcolor} % For cell coloring
\usepackage{booktabs}
\usepackage{multirow}
\usepackage{rotating}
\WsConferencePaper

\usepackage[T1]{fontenc}
\usepackage{dfadobe}  

\usepackage{cite}  % comment out for biblatex with backend=biber
\BibtexOrBiblatex
\electronicVersion
\PrintedOrElectronic
\ifpdf \usepackage{graphicx} \pdfcompresslevel=9
\else \usepackage[dvips]{graphicx} \fi

\usepackage{egweblnk} 
\title[PC-NCLaws: Physics-Embedded Conditional Neural Constitutive Laws for Elastoplastic Materials]%
      {PC-NCLaws: Physics-Embedded Conditional Neural Constitutive Laws for Elastoplastic Materials}

\author[Xueguang Xie \& Shu Yan et al.]
{\parbox{\textwidth}{\centering 
Xueguang Xie\thanks{joint first author; xgxie@ustb.edu.cn}$^{1}$,
Shu Yan\thanks{joint first author; shuyan03@u.nus.edu}$^{2}$,
Shiwen Jia$^{3}$,
Siyu Yang$^{1}$,
Aimin Hao$^{3}$\orcid{0000-0002-5774-6706},
Yang Gao$^{3}$\orcid{0000-0002-9149-3554},
Peng Yu\thanks{corresponding author; yupeng@buaa.edu.cn}$^{3}$\orcid{0000-0002-8652-2744}
        }
        \\
{\parbox{\textwidth}{\centering $^1$ School of Intelligence Science and Technology, University of Science and Technology Beijing, China\\
         $^2$School of Mathematical Sciences, Beihang University, Beijing\\
         $^3$State Key Laboratory of Virtual Reality Technology and Systems, Beihang University, China 
       }
}
}

\begin{document}

% uncomment for using teaser
% \teaser{
%  \includegraphics[width=1\linewidth]{figure/fig03.pdf}
%  \centering
%   \caption{
% Just by inputting the initial state of the object and the physical parameters of the groundtruth, our \textbf{PC-NClaws} can quickly generate a motion sequence, and it is close enough to the motion of the groundtruth. In particular, when the groundtruth changes, our model can generate the corresponding motion sequence without being retrained.}
% \label{fig03}
% }

\maketitle
%-------------------------------------------------------------------------

\begin{abstract}
While data-driven methods offer significant promise for modeling complex materials, they often face challenges in generalizing across diverse physical scenarios and maintaining physical consistency. To address these limitations, we propose a generalizable framework called Physics-Embedded Conditional Neural Constitutive Laws for Elastoplastic Materials, which combines the partial differential equations with neural networks. Specifically, the model employs two separate neural networks to model elastic and plastic constitutive laws. Simultaneously, the model incorporates physical parameters as conditional inputs and is trained on comprehensive datasets encompassing multiple scenarios with varying physical parameters, thereby enabling generalization across different properties without requiring retraining for each individual case. Furthermore, the differentiable architecture of our model, combined with its explicit parameter inputs, enables the inverse estimation of physical parameters from observed motion sequences. This capability extends our framework to objects with unknown or unmeasured properties. Experimental results demonstrate state-of-the-art performance in motion reconstruction, robust long-term prediction, geometry generalization, and precise parameters estimation for elastoplastic materials, highlighting its versatility as a unified simulator and inverse analysis tool.

%-------------------------------------------------------------------------
%  ACM CCS 1998
%  (see https://www.acm.org/publications/computing-classification-system/1998)
% \begin{classification} % according to https://www.acm.org/publications/computing-classification-system/1998
% \CCScat{Computer Graphics}{I.3.3}{Picture/Image Generation}{Line and curve generation}
% \end{classification}
%-------------------------------------------------------------------------
%  ACM CCS 2012
%The tool at \url{http://dl.acm.org/ccs.cfm} can be used to generate
% CCS codes.
%Example:


\end{abstract}

%-------------------------------------------------------------------------
\section{Introduction}
Traditional simulation methods such as FEM (Finite Element Method)\cite{hughes2003finite} and MPM (Material Point Method)\cite{jiang2016material,sulsky1995application} are driven by partial differential equations (PDEs). These approaches offer strong physical fidelity and are widely used across various domains. However, the constitutive laws in such methods are explicitly modeled, which can make it difficult to capture complex material behaviors. In recent years, with the rapid development of machine learning and deep learning, researchers begin to explore combining learning-based methods with traditional simulation techniques to enhance their overall capabilities. Some approaches, such as the Graph Network Simulator (GNS)\cite{sanchez2020learning}, replace the PDEs entirely with purely data-driven neural networks, which do not incorporate any known PDEs. While promising in flexibility, they can suffer from significant issues due to the lack of physical constraints, potentially resulting in physically incorrect modeling.

To address this, other approaches adopt a partial replacement strategy. NCLaws~\cite{ma2023learning} is a representative method. It argues that constitutive laws, due to their inherent complexity and variability, are well-suited to be learned from data, whereas the PDE-driven parts of the system are retained to ensure physical correctness. It is capable of accurately reconstructing the dynamics of the target object, performing long-term predictions, and generalizing to more complex geometries. However, the algorithm can only learn the motion of a single object with specific physical properties (e.g., fixed material physical parameters), and its evaluation experiments are entirely based on that particular object. when encountering new objects, the model's predictions tend to converge to similar outputs. In other words, the learned model cannot be transferred to predict the motion of objects with different physical parameters. To simulate the motion of a different object, the entire model must be retrained, resulting in significant computational inefficiency. Similarly, the GNS method also performs poorly in various generalization tasks. In light of these limitations, can we design a model capable of simultaneously learning the motion sequences of multiple objects with different physical parameters, thereby achieving motion sequence prediction without requiring individual object-specific training each time?
    
%     \item What novel significance would such capabilities confer upon the model? For instance, while maintaining high-fidelity simulation capacity, it could provide new perspectives for inverse problem research—such as inversely estimating physical parameters from motion data.%%%%
    
% \end{enumerate}

To solve the above problems, we propose "Physics-Embedded Conditional Neural Constitutive Laws (PC-NCLaws)", a novel hybrid simulation framework that combines neural networks with partial differential equations (hybrid NN-PDE). Inspired by the NCLaws method, we employ two separate neural networks to model elastic and plastic constitutive laws, respectively. Notably, our method introduces a key innovation by incorporating physical parameters related to material deformation as inputs to the two neural networks. The proposed model can learn the objects' dynamics with different physical attributes for specific materials through a single training process. During application, it is capable of rapidly inferring the motion of objects with parameters values outside the training datasets by simply providing the initial particle states and physical parameters without the need for retraining on each specific object, which greatly improves simulation efficiency as well as maintaining high-fidelity capacity. Additionally, for cases with unknown physical parameters, our method leverages the model’s robust forward inference capability and the neural network’s inherent differentiability, offering novel insights for inverse problem research, i.e. the estimation of physical parameters directly from motion data. Our model demonstrates excellent performance in multiple tasks across a range of elastoplastic materials, including elasticity, sand, and plasticine. In this paper, the "multiple scenarios" refer to multiple objects with the same material type but different physical parameters.

In summary, our work makes the following contributions:
\begin{itemize}
    \item We propose generalizable PC-NCLaws. It combines the partial differential equations with two separate neural networks to model elastic and plastic constitutive laws, which incorporate physical parameters as conditional inputs. Our method achieves both high-fidelity simulation and strong generalization across diverse physical parameters.
    \item We establish a novel inverse modeling approach based on our PC-NCLaws framework that enables accurate physical parameters estimation from motion observations, overcoming traditional ill-posedness challenges in properties identification. This breakthrough significantly extends our method's applicability to scenarios with unknown physical parameters, thereby substantially enhancing its generalization capability.  
    \item Our framework introduces a unified architecture for multi-material elastoplastic modeling (elasticity, sand, and plasticine), which eliminates material-specific structural adaptations while demonstrating consistent high-fidelity performance across all material categories.

\end{itemize}

\begin{figure*}[ht]
    \centering
    \includegraphics[width=1\linewidth]{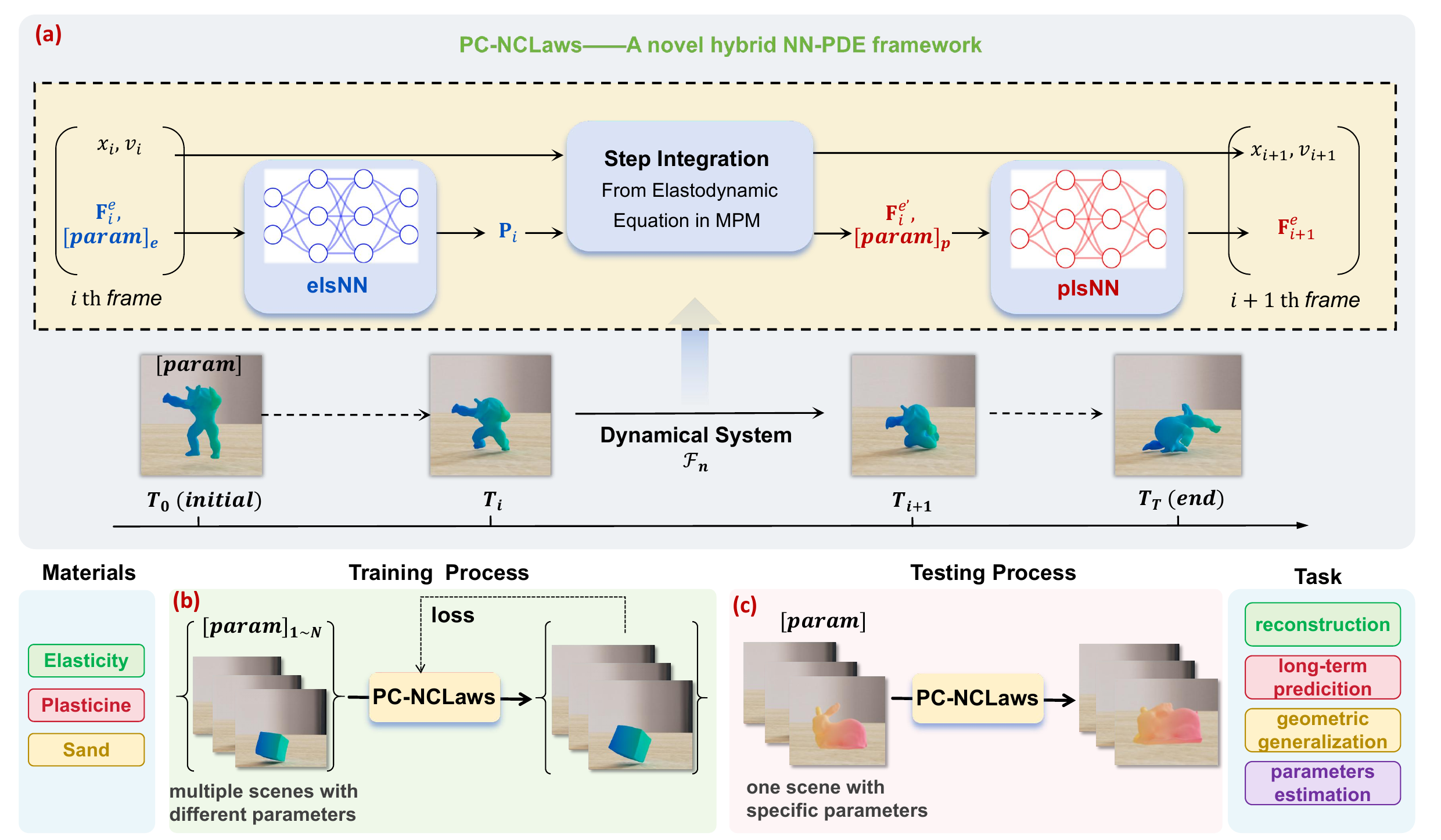}
    \caption{The 
\textbf{Pipeline} of our PC-NCLaws. We embedded two neural networks in the dynamical system to express the elastic and plastic constitutive law, respectively. At the \textbf{top row (a)}, we introduced physical parameters as inputs to guide the progress of motion. The bottom row shows the applicable materials of our model, the training process, and the testing process. In the \textbf{training process (b)}, we use multiple objects with the same material type but different physical parameters to train PC-NCLaws, and optimize the weights of the neural networks throughout the training. In the \textbf{testing process (c)}, the trained PC-NCLaws model acts as a high-fidelity simulator capable of performing various tasks.}
    \label{pipeline}
\end{figure*}

\section{Related Work}

\subsection{Constitutive Laws}
Constitutive laws are mathematical equations that reflect a material's response to external mechanical loads, describing the relationship between two physical quantities: such as stress and strain~\cite{fung1977first}, shear stress and shear rate~\cite{tropea2007springer}, and the electric displacement field and electric field~\cite{landau2013electrodynamics}. Constitutive laws play a key role in important partial differential equations such as the elastodynamic equations and Navier-Stokes equations. Constitutive laws have become fundamental in various branches of physics, such as elasticity~\cite{treloar1943elasticity, fung1967elasticity}, plasticity~\cite{drucker1952soil,mises1913mechanik}, and fluid mechanics~\cite{chhabra2023bubbles}. These traditional models are typically manually constructed from experimental observations and physical priors, often using high-order polynomials~\cite{ogden1997non} or spline functions~\cite{xu2015nonlinear}.

In recent years, inspired by the powerful fitting capabilities of neural networks, researchers have begun exploring data-driven neural network constitutive models~\cite{tartakovsky2018learning,shen2005finite,vlassis2020geometric,fuchs2021dnn2,klein2022polyconvex}. Their accuracy has surpassed traditional methods in various scenarios. These approaches enable unified representation of complex material behaviors and remove the need for explicit functional forms. Despite their promise, neural network constitutive models face challenges such as high data dependency, limited interpretability, and difficulty in maintaining physical consistency. To address this, researchers have proposed physically-constrained architectures and hybrid models that integrate neural networks with traditional formulations.

\subsection{Hybrid Neural Networks and PDE Dynamics}
Traditional numerical methods~\cite{jiang2016material} for solving partial differential equations (PDEs) face significant challenges in high-fidelity modeling of complex constitutive relations, while often relying on empirical parametric models. With recent advances in deep learning, researchers have developed various neural network-based approaches to enhance PDE solving. Physics-informed neural networks (PINNs)~\cite{raissi2019physics} incorporate PDE residuals as loss terms to enable mesh-free solutions, while operator learning methods like DeepONet~\cite{lu2019deeponet} and neural operators establish efficient surrogate mappings. Graph neural networks~\cite{sanchez2020learning} have shown particular promise by directly learning particle system dynamics while implicitly unifying constitutive relations and equations of motion. However, this algorithm demands extensive training data for effective learning. 

Additionally, there are many approaches that try to replace only part of the PDE; some choose to replace the equation of motion, while others try to use neural networks to learn the constitutive laws. Hybrid strategies such as data-driven discretization~\cite{bar2019learning} and APHYNITY framework~\cite{yin2021augmenting} combine physical principles with neural networks, and solver-in-the-loop~\cite{um2020solver} approaches integrate neural corrections into traditional iterations. PlasticityNet~\cite{NEURIPS2022_b235f0b4} replaces non-integrable elastoplastic forces with the gradient of a learned local potential energy. Rather than using analytical energy expressions, it employs a neural network to approximate forces via energy gradients, enabling stable and efficient simulation of plasticity—even in complex cases. While PlasticityNet generalizes across different Young’s modulus without retraining, it still requires retraining for new Poisson’s ratios. NCLaws~\cite{ma2023learning} replaces hand-crafted constitutive models with a neural network trained directly on motion data. Integrated into a PDE-based simulator, it preserves physical consistency and accurately models complex materials. Although NCLaws shows strong potential and serves as the foundation of our PC-NCLaws, its generalization is limited: it requires retraining for objects with unseen physical parameters—a major practical constraint. In summary, a common issue in these works is their limited generalizability or low efficiency.

\subsection{Estimation of physical parameters}
Estimating physical parameters of deformable objects from visual observations (e.g., motion videos) remains a long-standing challenge in computer graphics. Current research~\cite{ma2022risp,jatavallabhula2021gradsim,jaques2019physics,xie2024fluid} on identifying key physical parameters like Young's modulus, friction coefficients, and Poisson's ratio is still in its early stages. The parameters estimation task becomes particularly complex for deformable objects due to their high-dimensional state space and large deformation characteristics. Traditional approaches based on RGB-D sensors~\cite{bruns2023rgb} combined with physics simulation represent the most established non-data-driven solution, enabling interactive parameter recovery in virtual environments by tracking object deformation and elasticity in real-time. While widely adopted in VR/AR applications, these methods often require oversimplified constitutive models to maintain interactivity, leading to simulation inaccuracies and reduced estimation precision. 

Recent advances in deep learning enable differentiable optimization and rich representation learning, supporting data-driven constitutive models for high-quality simulation and parameter estimation. However, purely neural methods~\cite{xu2019densephysnet,li2018learning} often lack interpretability and fail to provide explicit physical parameters due to their black-box nature. Meanwhile, since inverse estimation of physical parameters typically involves minimizing the discrepancy between reconstructed and groundtruth, researchers have focused on improving reconstruction and rendering quality, as advancements in these modules can significantly enhance parameters estimation accuracy. PAC-NeRF~\cite{li2023pac} has introduced innovative approaches combining neural radiance fields~\cite{mildenhall2021nerf} with material point methods, enforcing conservation laws of continuum mechanics while estimating deformation parameters. Building upon this foundation, subsequent research PAC-NeRF+LPO~\cite{kaneko2024improving} developed Lagrangian particle optimization techniques that refine particle positions and features in Lagrangian space. These methods not only improve geometric accuracy but also boost the precision of physical parameters identification, particularly under sparse-view conditions. However, both approaches rely on predefined constitutive models, which may limit their performance in complex scenarios.

%-------------------------------------------------------------------------
\section{Method}
Our goal is to develop a novel hybrid NN-PDE framework, PC-NCLaws, capable of simultaneously learning and inferring constitutive laws for specific materials with diverse physical parameters. Section~\ref{Pre} analyzes conventional baseline models and their inability to handle multi-property learning. Building upon this analysis, Section~\ref{sec:PCNCLaws} presents our PC-NCLaws architecture in detail. Finally, Section~\ref{sec:paramest} derives the inverse problem algorithms for physical parameters estimation with this architecture.

\subsection{Preliminaries}\label{Pre}
To be self-contained, we present the foundational knowledge necessary to establish the basis for our model and algorithm.

\textbf{Elastodynamic Equation:} First, we assume that all materials in our work obey the elastodynamic equation~\cite{fung1977first}:
\begin{equation}
    \rho\,\ddot{\varphi}=\nabla\cdot \textbf{P}+\rho\, \textbf{b},
    \label{eq1}
\end{equation}
where $\rho$ is the initial density, $\varphi$ is the deformation map which describes the current position $\textbf{x}$ of an arbitrary point with an initial position $\textbf{X}$ and the time $t$, i.e., $\textbf{x}=\varphi(\textbf{X},t)$. Additionally, \textbf{P} is the first Piola-Kirchhoff stress, a function of the elastic part of the deformation gradient $\textbf{F}^e$ ("$e$" means the elastic): $\textbf{P}=p(\textbf{F}^e)$. $\textbf{b}$ is the body force, $\dot{\varphi}$ and $\ddot{\varphi}$ represent velocity and acceleration, respectively.

\textbf{Constitutive Laws:} In our work, we stipulate two constitutive laws. The elastic constitutive law defines the functional relationship between the stress $\textbf{P}$ and the elastic part of the deformation gradient $\textbf{F}^e$. The constitutive law of plasticity is an inequality constraint: $f_{\textit{Y}}(\textbf{P})<0$, where $f_{\textit{Y}}$ is the yield function. When $f_{\textit{Y}}=0$, the material will undergo plastic deformation. Constitutive laws, as defined above, are generalizable and can describe a wide range of materials.

\textbf{Dynamical System:}\, the simulation of an object's motion is driven by a dynamical system $\mathcal{F}_n\,(n=0,1,\cdots,T)$:
\begin{equation}
    s_{n+1}=\mathcal{F}_n(s_n),\, \forall n=0,1,\cdots,T,
    \label{eq2}
\end{equation}
where $T$ is the number of total time steps and  the state vector $s_i=\{\textbf{x}_i,\textbf{v}_i,\textbf{F}^e_i\}$ representing the physical state of the system at the $i$-th time step. In order to obtain the dynamical system $\mathcal{F}_n$, we can choose any numerical method, e.g., finite element methods, finite volume methods. In our work, we adopt the  MPM method~\cite{jiang2016material,sulsky1995application}. MPM is a hybrid Lagrangian-Eulerian computational technique that simulates continuum mechanics by tracking material points through a background grid. MPM discretizes the object with $M$ material points and $\mathcal{F}_n$ can be represented in Algorithm~\ref{al 1}.
The elastic constitutive law \textbf{els} computes the first Piola-Kirchhoff stresses $\textbf{P}_i$ through a traditional elastic model, e.g., the Saint Venant-Kirchhoff (StVK)~\cite{saint1856memoire}. The plastic constitutive laws \textbf{pls} project $\textbf{F}_{i+1}^{e'}$ to $\textbf{F}_{i+1}^{e}$ for plasticity constraints. The step integration $\mathcal{I}_i$ is derived from Eqn.~\ref{eq1} and follows the standard MPM process. This step integration is employed to solve the elastodynamic equation and update material points information.

\begin{algorithm}[htbp]
\caption{$\mathcal{F}_n$ (MPM)}
\label{al 1}
\begin{algorithmic}[1]
\State \textbf{Input:} Current state $s_i = \{\textbf{x}_i, \textbf{v}_i, \mathbf{F}_i^e\}$
\State \textbf{Output:} Next state $s_{i+1} = \{\textbf{x}_{i+1}, \textbf{v}_{i+1}, \textbf{F}_{i+1}^e\}$
\For{$j = 1$ to $M$}
    \State $\textbf{P}_i(j) \gets \textbf{els} \big(\textbf{F}_i^e(j)\big)$
\EndFor
\State $\{\textbf{x}_{i+1}, \textbf{v}_{i+1}, \textbf{F}_{i+1}^{e'}\} \gets \mathcal{I}_i(\textbf{x}_i, \textbf{v}_i, \textbf{F}_i^e, \textbf{P}_i)$
\For{$j = 1$ to $M$}
    \State $\textbf{F}_{i+1}^e(j) \gets \textbf{pls} \big(\textbf{F}_{i+1}^{e'}(j)\big)$
\EndFor
\end{algorithmic}
\end{algorithm}
The current hybrid NN-PDE model NCLaws~\cite{ma2023learning} replaces the $\textbf{els}$ and $\textbf{pls}$ in Algorithm~\ref{al 1} with two neural networks: $\textbf{elsNN}$ and $\textbf{plsNN}$, which input the elastic part of the deformation gradient $\textbf{F}^e$ at the material point at that moment in time, and output stresses $\textbf{P}_i$ and  $\textbf{F}_{i+1}^{e}$, respectively. This method can effectively learn the elastic and plastic constitutive laws of a single object with specific physical parameters, which maintain physical correctness through known PDE and enhance the model's ability to model the complex nonlinear behavior of materials. However, it is difficult to learn the constitutive laws of objects with different physical parameters because this network does not incorporate physical parameters as input labels. Therefore, it is essential to develop a novel hybrid NN-PDE dynamical system with generalization capabilities that can correspondingly adapt to objects with different physical parameters.

\subsection{Physics-Embedded Conditional Neural Constitutive Laws }\label{sec:PCNCLaws}

\subsubsection{PC-NCLaws framework}
To address the aforementioned issues, we propose a universal NN-PDE framework called PC-NCLaws, capable of simultaneously learning and inferring constitutive laws for specific materials with diverse physical parameters objects. The algorithmic framework is illustrated in Figure~\ref{pipeline}(a). First, we embed two neural networks within the dynamical PDE system to represent the elastic and plastic constitutive laws, respectively. Simultaneously, we incorporate physical parameters as conditional inputs to the neural networks to guide the motion process, ensuring its generalizability. Our physics-embedding neural framework as detailed in Algorithm~\ref{al 3} and Figure~\ref{pipeline}(a), advances material states through a three-stage computational pipeline. The input to our algorithm includes the current state $\{\textbf{x}_{i}, \textbf{v}_{i}, \textbf{F}_{i}^e\}$, physical parameters related to elasticity $[param]e$ (e.g., Young's modulus and Poisson's ratio), and physical parameters related to plasticity $[param]p$ (e.g., yield stress). The output is $\{\textbf{x}_{i+1}, \textbf{v}_{i+1}, \textbf{F}_{i+1}^e\}$.

\begin{algorithm}[htbp]
\caption{$\mathcal{F}_n$ (Ours)}
\label{al 3}
\begin{algorithmic}[1]
\State \textbf{Input:} Current state $s_i = \{\textbf{x}_i, \textbf{v}_i, \mathbf{F}_i^e\}$,\\ ~~~~~~~~~~~~physical parameters related to elasticity $[param]_e$ , \\~~~~~~~~~~~~physical parameters related to plasticity $[param]_p$
\State \textbf{Output:} Next state $s_{i+1} = \{\textbf{x}_{i+1}, \textbf{v}_{i+1}, \textbf{F}_{i+1}^e\}$
\For{$j = 1$ to $M$}
    \State $\textbf{P}_i(j) \gets \textcolor{blue}{\textbf{elsNN}} \big(\textbf{F}_i^e(j), \textcolor{blue}{[param]_e}\big)$
\EndFor
\State \#{Step Integration from the elastodynamic equation in MPM}
\State $\{\textbf{x}_{i+1}, \textbf{v}_{i+1}, \textbf{F}_{i+1}^{e'}\} \gets \mathcal{I}_i(\textbf{x}_i, \textbf{v}_i, \textbf{F}_i^e, \textbf{P}_i)$ 
\For{$j = 1$ to $M$}
    \State $\textbf{F}_{i+1}^e(j) \gets \textcolor{red}{\textbf{plsNN}} \big(\textbf{F}_{i+1}^{e'}(j),\textcolor{red}{[param]_p}\big)$
\EndFor
\end{algorithmic}
\end{algorithm}

\begin{figure*}[htbp]
    \centering
    \includegraphics[width=0.85\linewidth]{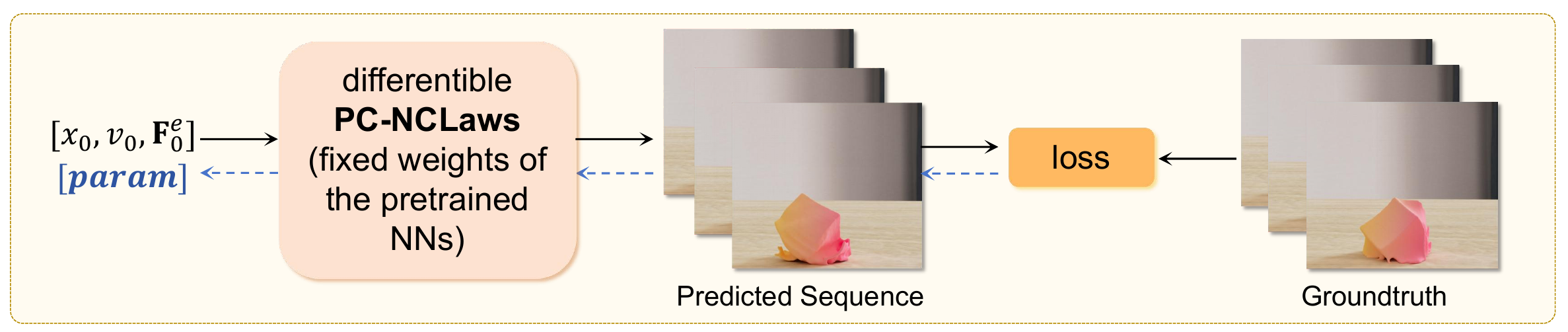}
    \caption{\textbf{Parameters Estimation.} After PC-NCLaws is trained, we can use it to reversely estimate the physical parameters from the object's motion sequence. By initializing a set of physical parameters and other physical information, we use PC-NCLaws for simulation, compute the loss with the ground truth, and backward the gradient to optimize the physical parameters. During this process, the weights of the neural networks in PC-NCLaws remain unchanged. When the loss converges, we consider the physical parameters at this time to be the properties of this object.}
    \label{pipeline2}
\end{figure*}

The computation begins with the elastic phase where the blue-colored elastic neural network (\textcolor{blue}{elsNN}) processes each material point's deformation gradient (elastic part) $\textbf{F}_i^e(j)$ along with elastic parameters $[param]_e$ to compute the first Piola-Kirchhoff stress $\textbf{P}_i(j)$. These stresses then drive the MPM integration step $\mathcal{I}_i$, which updates the particle positions $\textbf{x}_{i+1}$, velocities $\textbf{v}_{i+1}$, and produces an intermediate part of the deformation gradient $\textbf{F}_{i+1}^{e'}$. In the subsequent plastic phase, the red-colored plastic neural network (\textcolor{red}{plsNN}) processes this intermediate deformation gradient along with plastic parameters $[param]_p$ to compute the final deformation gradient $\textbf{F}_{i+1}^e(j)$, effectively modeling plastic flow and hardening effects. This decoupled architecture allows separate neural representations of elastic and plastic responses while maintaining exact momentum conservation through the physics-constrained MPM integration, with both neural network evaluations operating efficiently across all materials with different physical parameters.

Notably, when trained on the constitutive law from a single simulation scenario (mentioned as \textbf{PC-NCLaws-S} in Section~\ref{exp}), our model demonstrates similarities to the hybrid framework described in Section~\ref{Pre}. However, our approach achieves substantial improvements in both task versatility and generalization performance. This enhanced capability stems from our physics-embedded conditional learning paradigm that explicitly encodes material parameters as network inputs rather than learned constants.

\subsubsection{Network Training}

During the training process, as shown in Figure~\ref{pipeline}(b), we use multiple scenarios featuring the same kind of material with different physical parameters as the dataset to train PC-NCLaws. Our framework can handle various kind of materials, including elasticity, plasticine, and sand, though each material requires independent training due to their inherently distinct constitutive laws.

The neural networks are trained by minimizing a position-based loss function that measures the discrepancy between predicted particle trajectories and ground truth data by the mean squared error (MSE). For a epoch of $K$ objects simulated over $T$ time steps, the loss function is defined as:
\begin{equation}
\mathcal{L}_{\text{total}}(\theta_e, \theta_p) = \sum_{k=1}^K (\frac{1}{T}\sum_{n=0}^T \Vert \mathbf{x}_{k,n}(\theta_e,\theta_p) - \mathbf{x}_{k,n}^{\text{gt}} \Vert_2^2),
\label{eq:loss_function}
\end{equation}
where $\theta_e$ and $\theta_p$ are the trainable parameters of the elastic network (elsNN) and plastic network (plsNN), respectively. $\mathbf{x}_{k,n}$ and $\mathbf{x}_{k,n}^{\text{gt}}$ represent the predicted and ground-truth particle positions for the $k$-th object at time step $n$. $\Vert\cdot\Vert_2$ denotes the Euclidean norm (L2 distance). The double summation accumulates errors across all objects ($K$) and all time steps ($T$) in each training epoch.

\subsubsection{Simulation Testing}

In the testing phase as shown in Figure~\ref{pipeline}(c), PC-NCLaws serves as a high-fidelity simulator capable of accomplishing a series of tasks, including reconstruction, long-term prediction, generalization to different geometric objects, and parameters estimation. Notably, our framework directly adapts to objects with varying physical parameters without requiring retraining. Another key highlight is parameters estimation, an exclusive inverse modeling capability of our approach, which will be detailed in the next section. This inverse modeling capability is a key differentiator from other NN-PDE simulators like NCLaws or GNS.

\subsection{physical parameters Estimation from Motion Sequence}\label{sec:paramest}

Leveraging the differentiable nature of our neural network and its explicit parameter input architecture, we employ the trained PC-NCLaws model to estimate physical parameters from a motion sequence $\{\mathbf{x}_m^{\text{gt}}\}_{m=0}^{T'}$ ($T'$ is the observation window length) of an object with known material type but unknown parameters. This capability enables our model to support various downstream tasks involving objects with unknown physical parameters through the estimated parameters. As detailed in Figure~\ref{pipeline2}, the estimation process maintains fixed neural network weights ($\theta_e,\theta_p$ frozen) while optimizing $[param]$ via gradient descent (Adam/SGD) through the differentiable PC-NCLaws model. The converged parameters minimize the trajectory discrepancy, yielding the estimated material properties. We formulate the estimation task through the following position-matching loss:
\begin{equation}
\mathcal{L}_{\text{p}}([param]) = \frac{1}{T'}\sum_{m=0}^{T'} \Vert\mathbf{x}_m([param]) - \mathbf{x}_m^{\text{gt}}\Vert_2^2,
\label{eq:inverse_loss}
\end{equation}
where $[param]$ denotes the trainable physical parameters (elastic/plastic coefficients), $\mathbf{x}_m([param])$ represents the simulated positions using current parameters.
\begin{table*}[htbp]
\centering
\renewcommand{\arraystretch}{1}
\caption{\textbf{Comparison.} We present quantitative results between our method and baselines on three materials. Notably, since the \textbf{Splines} method uses spline curves to represent material properties, both \textbf{GNS} and \textbf{NCLaws} are designed to learn motion for single object, while our model typically handles multiple scenarios, a direct comparison is not straightforward. To address this, we specifically adapt our model to learn single-object motion (\textbf{PC-NCLaws-S}) for fair benchmarking against baselines. Additionally, we demonstrate our model's performance on single objects after being trained on multiple-object motion (\textbf{PC-NCLaws-M}).}
\label{comparison}
\begin{tabular}{cccccccccc}
\toprule
\multirow{2}{*}{\textbf{Method}} & \multicolumn{3}{c}{\textbf{Elasticity}} & \multicolumn{3}{c}{\textbf{Plasticine}} & \multicolumn{3}{c}{\textbf{Sand}}\\
\cmidrule(lr){2-10}
&\textbf{Recon} & \textbf{Pred} & \textbf{Geo} & \textbf{Recon} & \textbf{Pred} & \textbf{Geo} & \textbf{Recon} & \textbf{Pred} & \textbf{Geo}\\
\midrule
Spline & 2.4e-1 & 2.6e-1 & 3.5e-1 & 3.0e-1 & 3.0e-1 & 3.1e-1 & 3.2e-1 & 3.5e-1 & 3.6e-1 \\
GNS & 2.1e-2 & 3.3e-2 & 1.6e-1 & 8.7e-3 & 1.1e-2 & 3.7e-2 & 1.1e-2 & 1.5e-2 & 3.4e-1 \\
NCLaws & 2.4e-4 & 9.8e-4 & 4.1e-4 & 6.5e-5 & 1.4e-4 & 2.3e-4 & 2.6e-5 & 4.2e-5 & 3.6e-4 \\
\textbf{PC-NCLaws-S} & 5.9e-5 & 1.4e-4 & 3.2e-4 & 8.6e-5 & 2.6e-4 & 4.5e-4 & 5.8e-5 & 1.3e-4 & 3.3e-4\\

\textbf{PC-NCLaws-M} & 2.6e-4 & 2.2e-3 & 2.1e-3 & 4.1e-4 & 5.2e-3 & 4.0e-3 & 1.5e-4 & 1.1e-3 & 5.4e-4 \\
\bottomrule
\end{tabular}
\end{table*}

%-------------------------------------------------------------------------
\section{Experiments and Results}

\subsection{Experiment Setups}

 Our model supports a variety of materials, and in this work we validate its performance on three important elastoplastic materials: elasticity, sand, and plasticine. We use the fixed corotated hyperelastic material~\cite{stomakhin2012energetically} to model elasticity without plastic components. For sand, where frictional behavior dominates plasticity, we employ the St. Venant-Kirchhoff (StVK) model~\cite{saint1856memoire} for elasticity combined with the Drucker-Prager yield condition~\cite{drucker1952soil} for plasticity. Plasticine, characterized by cohesive-force-dominated plasticity, also uses the StVK model for elasticity but adopts the von Mises model~\cite{mises1913mechanik} for its plastic behavior. The key physical parameters governing these materials' dynamic behaviors are:
\begin{enumerate}
\item \textit{Elasticity}: Young's modulus $E$ (material stiffness), Poisson's ratio $\nu$ (volumetric deformation resistance)
\item \textit{Sand}: Young's modulus $E$, Poisson's ratio $\nu$, and friction angle $\theta_f$ (internal friction coefficient)
\item \textit{Plasticine}: Young's modulus $E$, Poisson's ratio $\nu$, and yield stress $\tau_Y$ (permanent deformation threshold)
\end{enumerate}

We establish physically plausible parameter ranges for each material based on fundamental physical laws and empirical observations. For each material type, we generate training data by simulating multiple distinct scenes with parameter values sampled across these predetermined ranges (more than 10 configurations per material) . Each configuration produces a 500-timestep motion sequence capturing the object's gravitational free-fall and subsequent ground collision dynamics. The objects are cubic volumes uniformly discretized into 1,000 material points. Importantly, we adopt a cross-scenario partitioning scheme for our training and test sets, ensuring that all scenarios used in testing tasks contain physical parameters strictly excluded from the training data - meaning the network encounters entirely novel parameter configurations during evaluation.

Our neural network architecture employs a standard 3-layer multilayer perceptron (MLP) with 64 neurons in each hidden layer. We train the model using the Adam optimization algorithm with an initial learning rate that follows a cosine annealing schedule during training. All experiments are conducted on an NVIDIA RTX 4090 GPU.

\subsection{Reconstrution and Generalization} \label{exp}

\begin{figure*}
    \centering
    \includegraphics[width=1\linewidth]{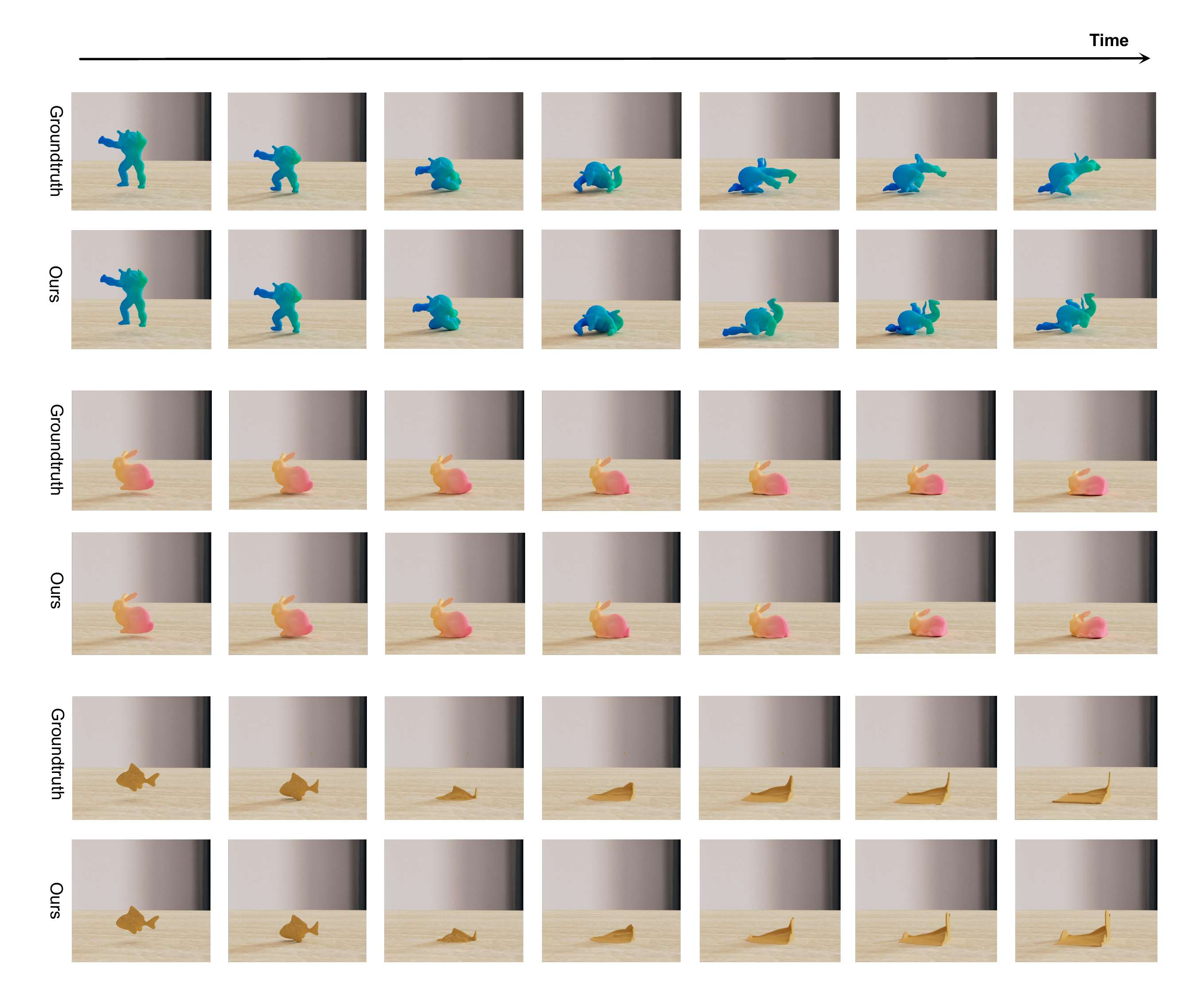}    \caption{\textbf{Qualitative Results} (via \textbf{PC-NCLaws-M} training strategy). We present visual results for the complex geometric generalization.}
    \label{geo}
\end{figure*}

In this subsection, we evaluate our model's performance across four key tasks: (1) \textbf{Reconstruction} - accurately reproducing observed motion sequences from input dynamics; (2) \textbf{Long-Term Prediction} - forecasting object trajectories over double extended time horizons beyond the training window; (3) \textbf{Geometric Generalization} - adapting to objects with novel, more complex geometries not encountered during training; and (4) \textbf{physical parameters Generalization} - predicting motions for objects with physical parameters outside the training distribution. This comprehensive evaluation tests both the model's fidelity in reconstructing known behaviors and its generalization capabilities across temporal, geometric, and physical domains.

To further demonstrate the superiority of our approach, we select three learning-based methods as comparative baselines. The \textbf{Spline}\cite{xu2015nonlinear} model utilizes B\'ezier splines to represent the constitutive laws of objects and optimizes the spline parameters during training. \textbf{GNS} (Graph Network Simulator)~\cite{sanchez2020learning} adopts graph neural networks to learn particle dynamics without incorporating PDE constraints. \textbf{NCLaws}~\cite{ma2023learning} learns both elastic and plastic constitutive laws of a single scene while being driven by PDE. We present the comparative results of the first three experiments in Table~\ref{comparison}, where the quantitative results report MSE averaged every 5 time steps.

Since the methods in the baselines cannot train multiple scenes as our method, a direct comparison may be less persuasive. In this regard, we adopt two different training strategies for our method PC-NCLaws, namely \textbf{PC-NCLaws-S}(single) and \textbf{PC-NCLaws-M}(multiple). The former is trained with one single scene with an object and can be directly compared with the existing baselines. The latter is trained with multiple scenarios to verify the generalization ability and functional diversity of our model. Notably, while the three baseline methods and \textbf{PC-NCLaws-S} were all trained on a single scenario, the experimental setups for it are kept the same as those of the baselines.

\begin{table*}[ht]
\centering
\small
\renewcommand{\arraystretch}{1.2}
\caption{\textbf{Parameters Estimation Results.} Quantitative evaluation of physical parameters estimation for objects with unknown properties. PC-NCLaws achieves high-accuracy predictions across all materials, with test objects excluded from training to ensure validity.}
\begin{tabular}{c c c c }
\hline
\textbf{Material} & \textbf{Initial Guess} & \textbf{Result} & \textbf{Ground Truth} \\
\hline
\multirow{3}{*}{Elasticity} 
& $E = 5 \times 10^4$, $\nu = 0.02$ & $E = 1.84 \times 10^5$, $\nu = 0.20$ & $E = 1.80 \times 10^5$, $\nu = 0.13$ \\
& $E = 5 \times 10^4$, $\nu = 0.5$ & $E = 1.50 \times 10^5$, $\nu = 0.22$ & $E = 1.40 \times 10^5$, $\nu = 0.18$ \\
& $E = 1 \times 10^5$, $\nu = 0.03$ & $E = 2.06 \times 10^5$, $\nu = 0.19$ & $E = 2.00 \times 10^5$, $\nu = 0.15$ \\

\multirow{3}{*}{Plasticine}
& $E = 5 \times 10^4$, $\nu = 0.10$, $\tau_Y = 1.0 \times 10^3$ & $E = 1.13 \times 10^5$, $\nu = 0.41$, $\tau_Y = 3.1 \times 10^3$ & $E = 4.70 \times 10^5$, $\nu = 0.36$, $\tau_Y = 4.2 \times 10^3$ \\
& $E = 1 \times 10^5$, $\nu = 0.10$, $\tau_Y = 1.5 \times 10^3$ & $E = 2.92 \times 10^5$, $\nu = 0.45$, $\tau_Y = 4.6 \times 10^3$ & $E = 5.3 \times 10^5$, $\nu = 0.31$, $\tau_Y = 5.1 \times 10^3$ \\
& $E = 1 \times 10^5$, $\nu = 0.05$, $\tau_Y = 2 \times 10^3$ & $E = 4.48 \times 10^5$, $\nu = 0.39$, $\tau_Y = 5.3 \times 10^3$ & $E = 4.2 \times 10^5$, $\nu = 0.33$, $\tau_Y = 5.7 \times 10^3$ \\

\multirow{3}{*}{Sand} 
& $E = 1 \times 10^5$, $\nu = 0.05$, $\theta_f = 10^\circ$ & $E = 3.74 \times 10^5$, $\nu = 0.19$, $\theta_f = 29.7^\circ$ & $E = 3.70 \times 10^5$, $\nu = 0.18$, $\theta_f = 30.0^\circ$ \\
& $E = 1 \times 10^5$, $\nu = 0.1$, $\theta_f = 8^\circ$ & $E = 3.36 \times 10^5$, $\nu = 0.21$, $\theta_f = 25.1^\circ$ & $E = 4.2 \times 10^5$, $\nu = 0.23$, $\theta_f = 26.0^\circ$ \\
& $E = 1 \times 10^5$, $\nu = 0.05$, $\theta_f = 5^\circ$ & $E = 2.71 \times 10^5$, $\nu = 0.27$, $\theta_f = 22.0^\circ$ & $E = 3.30 \times 10^5$, $\nu = 0.20$, $\theta_f = 22.0^\circ$ \\
\hline
\end{tabular}
\label{tab:parameter_estimation}
\end{table*}

Qualitative results for the first three tasks (via \textbf{PC-NCLaws-M} training strategy) are presented in Figure~\ref{geo}, demonstrating nearly identical alignment with groundtruth (GT) data. Meanwhile, the quantitative results as shown in Table~\ref{comparison} reveal \textbf{Spline} consistently underperforms due to optimization instability from discontinuous control points and the limited flexibility of quadratic splines. This is exacerbated by boundary effects where endpoint curvature cannot be properly constrained. \textbf{GNS} method delivers acceptable performance across all three materials, with loss magnitudes around 1e-3 to 1e-2, suggesting it has roughly learned the constitutive models. Sometimes it occasionally exhibits higher loss due to its lack of PDE constraints, resulting in physical inconsistencies. 
In comparison, both \textbf{PC-NCLaws-S} and \textbf{NCLaws} demonstrate robust performance, achieving consistent losses below 1e-3. While the two methods show comparable performance in most test cases, our model's explicit incorporation of physical parameters as inputs provides superior fitting of constitutive laws. When trained with identical strategies, our approach matches the state-of-the-art in accuracy. \textbf{PC-NCLaws-M} also demonstrates remarkable accuracy - while slightly inferior to \textbf{PC-NCLaws-S}, its predictions remain sufficiently close to ground truth, as evidenced by qualitative results. More importantly, the training paradigm of \textbf{PC-NCLaws-M} endows the model with significantly superior generalization capability compared to baseline methods. It maintains this level of accuracy when handling tasks involving objects with arbitrary physical parameters within reasonable ranges, whereas the baseline models typically only achieve high accuracy for the specifically trained objects.

The physical parameters generalization task demonstrates our model's most significant contribution, showcasing its ability to extrapolate beyond the training parameter distribution. Unlike the three baselines that require retraining when simulating a new object (e.g., NCLaws takes approximately 30 minutes to learn each new object's constitutive laws), our approach \textbf{PC-NCLaws-M} only needs to be trained once to learn constitutive laws for multiple scenarios with various physical parameters. By leveraging neural networks' inherent interpolation capabilities, our model can directly predict motion sequences for various objects without retraining - even for objects not included in the training set, it achieves accurate reconstruction in a few seconds, significantly improving both efficiency and generalization. Simply by inputting the initial particle states and the object's physical parameters, our model can generate the complete motion sequence. As shown in Table~\ref{tab:material_parameters}, our method maintains excellent precision (achieving losses on the order of 1e-4) when predicting motion sequences for unseen objects, demonstrating comparable accuracy to previous reconstruction experiments and proving its remarkable transfer learning capability.

\begin{table}[htbp]
\centering
\small
\renewcommand{\arraystretch}{1.2}
\caption{The motion sequence prediction for \textbf{unseen} objects can also be controlled at the order of magnitude of $10^{-4}$, demonstrating extremely high precision.}
\label{tab:material_parameters}
\begin{tabular}{ccc}
\hline
\textbf{Material} & \textbf{Object}&\textbf{Loss} \\
\hline
\multirow{2}{*}{Elasticity} & $E = 1.8 \times 10^5$, $\nu = 0.13$ & $\mathbf{4.3 \times 10^{-4}}$\\ 
           & $E = 1.4 \times 10^5$, $\nu = 0.18$ & $\mathbf{3.9 \times 10^{-4}}$\\

\multirow{2}{*}{Plasticine} & $E = 5.3 \times 10^5$, $\nu = 0.31$, $\tau_Y = 5.1 \times 10^3$ & $\mathbf{3.4 \times 10^{-4}}$\\
           & $E = 4.2 \times 10^5$, $\nu = 0.33$, $\tau_Y = 5.7 \times 10^3$ & $\mathbf{3.1 \times 10^{-4}}$\\

\multirow{2}{*}{Sand}       & $E = 3.7 \times 10^5$, $\nu = 0.18$, $\theta_f = 30^\circ$&  $\mathbf{2.0 \times 10^{-4}}$\\
           & $E = 4.2 \times 10^5$, $\nu = 0.23$, $\theta_f = 26^\circ$ &$\mathbf{2.6 \times 10^{-4}}$\\
\hline
\end{tabular}
\end{table}

% \begin{figure*}
%     \centering
%     \includegraphics[width=1\linewidth]{figure/fig01.pdf}
%     \caption{Reconstruction (Elasticity)}
%     \label{fig01}
% \end{figure*}

% \begin{figure*}
%     \centering
%     \includegraphics[width=1\linewidth]{figure/fig06.pdf}
%     \caption{Long-Term Prediction (Sand)}
%     \label{fig06}
% \end{figure*}

% \begin{figure*}
%     \centering
%     \includegraphics[width=1\linewidth]{figure/fig05.pdf}
%     \caption{Geometric Generalization (Plasticine)}
%     \label{fig05}
% \end{figure*}

In summary, our model demonstrates outstanding performance across all four aforementioned experiments, significantly outperforming both \textbf{splines} and \textbf{GNS} approaches in terms of both accuracy and generalizability. While showing slightly lower precision than \textbf{NCLaws} on single-scene tasks (with the difference being practically negligible), it exhibits overwhelming advantages in model transfer capability, which is a decisive edge that all three baselines fundamentally lack. We therefore assert that our method achieves an optimal balance between high-fidelity accuracy and exceptional generalization capability.

\begin{figure}[hbp]
  \centering
  \includegraphics[width=0.95\linewidth]{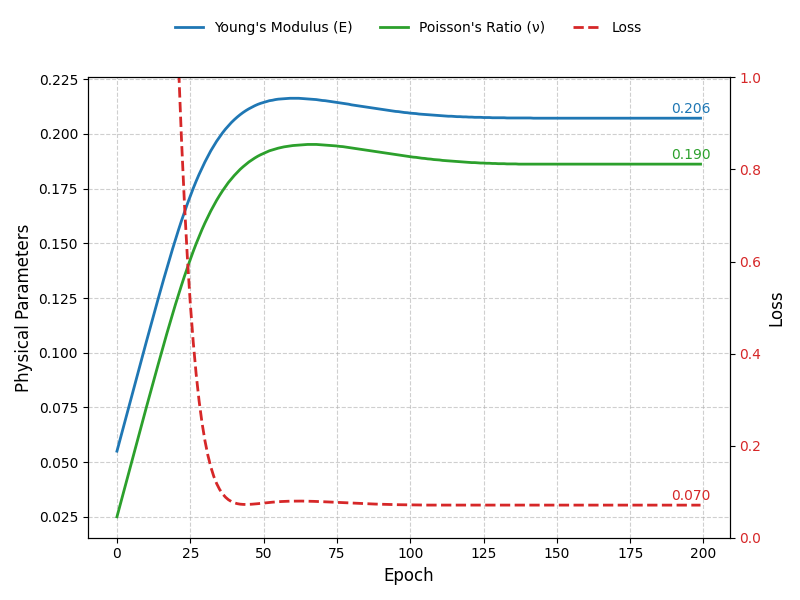}
  \caption{
  The optimization process for \textbf{Elasticity} parameters estimation (the third experiment set, groundtruth: $E = 2.00 \times 10^5$, $\nu = 0.15$). Young's modulus and Poisson's ratio evolution are shown with uniform scaling. The red curve tracks the error.}
\label{fig7}
\end{figure}

\begin{figure*}[t]
    \centering
    \includegraphics[width=0.9\linewidth]{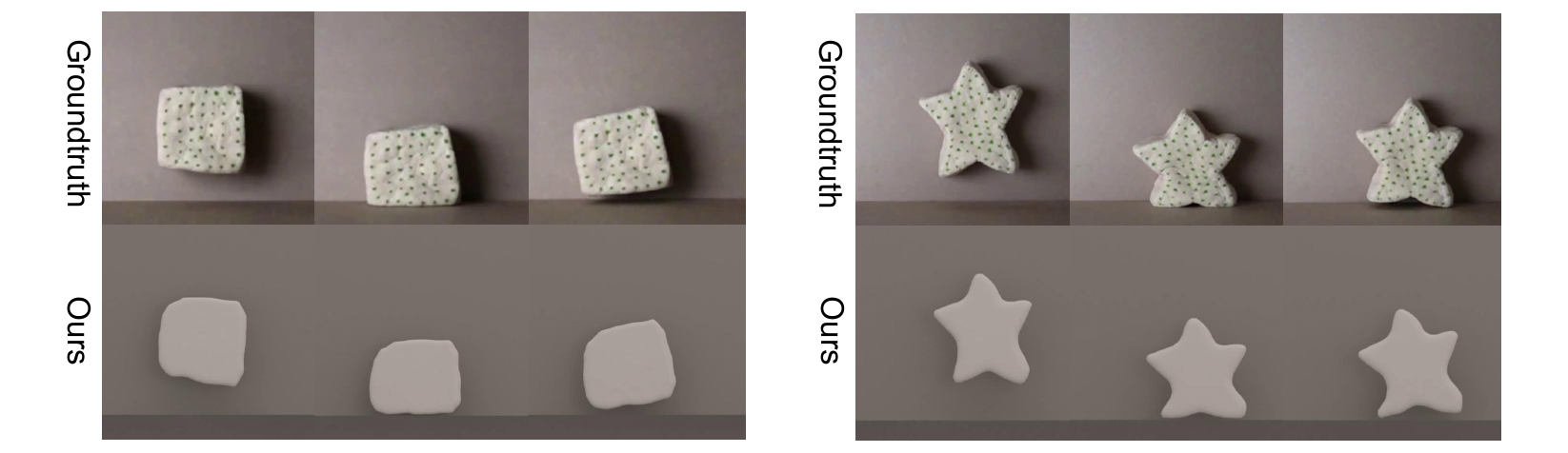}    \caption{\textbf{Real-World Experiments.} Two dough samples with distinct initial shapes (cube: simple; star: complex) were released from rest at a height. Physical parameters were inversely estimated from captured motion trajectories and used for forward simulation from identical initial conditions. Top: ground truth; bottom: corresponding simulation results.}
    \label{real}
\end{figure*}

\subsection{Parameters Estimation}

While existing learning-based models lack the capacity for explicit parameter conditioning - they neither generalize to scenarios with different physical parameters nor possess parameters estimation capabilities. In contrast, the differentiable nature of this framework and its explicit parameter input, our model can perform inverse modeling tasks, such as parameters estimation.

We conduct nine parameter estimation experiments (three per material), ensuring rigorous scientific evaluation by explicitly excluding all test objects from the training set. The quantitative results are presented in Table~\ref{tab:parameter_estimation}. Each row specifies the material, initial guess, optimized result, and groundtruth. The results demonstrate our model's exceptional performance in physical parameter estimation tasks, with optimized parameters closely matching the ground truth.

For \textbf{elasticity} and \textbf{sand}, our model achieves particularly accurate predictions of their physical parameters, with most results exhibiting errors below 5\%. Notably, for the friction angle in sand, the error remains within $1^\circ$ across all experiments - demonstrating exceptional precision. Estimating physical parameters for \textbf{plasticine} is particularly challenging due to its complex and strongly coupled physical properties, which often cause optimization methods to get stuck in local minima. While current state-of-the-art approaches sometimes can produce errors of up to an order of magnitude ($\sim 10\times$) in some parameters, our method consistently delivers predictions within physically plausible ranges, demonstrating robust performance on this difficult task.

\subsection{Real-World Experiements}

We further validate our model on real-world data using the dataset captured by NCLaws. The data consists of two distinct scenarios involving objects made of dough in cube and star shapes. In each scenario, a piece of dough is released from a height, falls, and rebounds from the ground. The deformation process is captured through tracked markers on the object’s surface and recorded for analysis. We used the deformation data from the cube-shaped sample to inversely estimate the physical parameters. These estimated parameters, along with the initial point cloud state, were then fed into our model to simulate the corresponding falling and rebounding process. The results are shown in Figure~\ref{real} (left). Our simulated motion trajectories and deformations align closely with the ground truth, demonstrating the strong generalization capability of our model on real-world data. Additionally, we performed a geometry generalization test with a complex star-shaped dough object. Using the same estimated parameters and initial point cloud, we simulated its dynamic behavior, as shown in Figure~\ref{real} (right). It confirm the model’s strong generalization to unseen geometries. Notably, unlike NCLaws, our approach requires no retraining of the constitutive model, greatly improving efficiency. The corresponding simulations are provided in the supplementary video.

\section{Conclusion}

We present PC-NCLaws, a generalizable framework for modeling elastoplastic materials by decoupling elastic and plastic constitutive laws into two separate neural networks conditioned on physical parameters. This architecture enables strong generalization across diverse scenarios without the need for material-specific retraining, which many existing methods require or fail to handle without losing physical consistency. The differentiable, parameter-conditioned design also supports inverse estimation of material parameters from observed motion, facilitating downstream tasks on objects with unknown properties. Experiments demonstrate superior performance in motion reconstruction, long-term stability, generalization to unseen geometries, and parameter inference accuracy. 

During training across multiple scenarios, the selection, range, and amount of data significantly impact performance: too much data increases cost, while too little weakens generalization. Parameter coupling can also cause local optima in estimation, especially for certain materials, posing a challenge for future work. Potential extensions include adapting PC-NCLaws to rigid bodies, fluids, and non-Newtonian materials, and improving its robustness for real-world deployment.

\section{Acknowledgement}
This work was supported in part by the National Key R\&D Program of China (No.2023YFC3604500), the Postdoctoral Fellowship Program of CPSF under grant number GZC-20233375, the Guangxi Science and Technology Major Program (No. GuiKeAA24206017), the Open Project Program of State Key Laboratory of Virtual Reality Technology and Systems, Beihang University (No.VRLAB2025C16), and the Beijing Hospitals Authority Clinical Medicine Development of special funding support, code: YGLX202525.

%-------------------------------------------------------------------------

%------------------------------------------------------------------------

%-------------------------------------------------------------------------
% bibtex
\bibliographystyle{eg-alpha-doi} 
\bibliography{egbibsample}       

% biblatex with biber
% \printbibliography                

%-------------------------------------------------------------------------
%Color tables are no longer required for purely electronic publications.

\begin{figure*}
    \centering
    \includegraphics[width=1\linewidth]{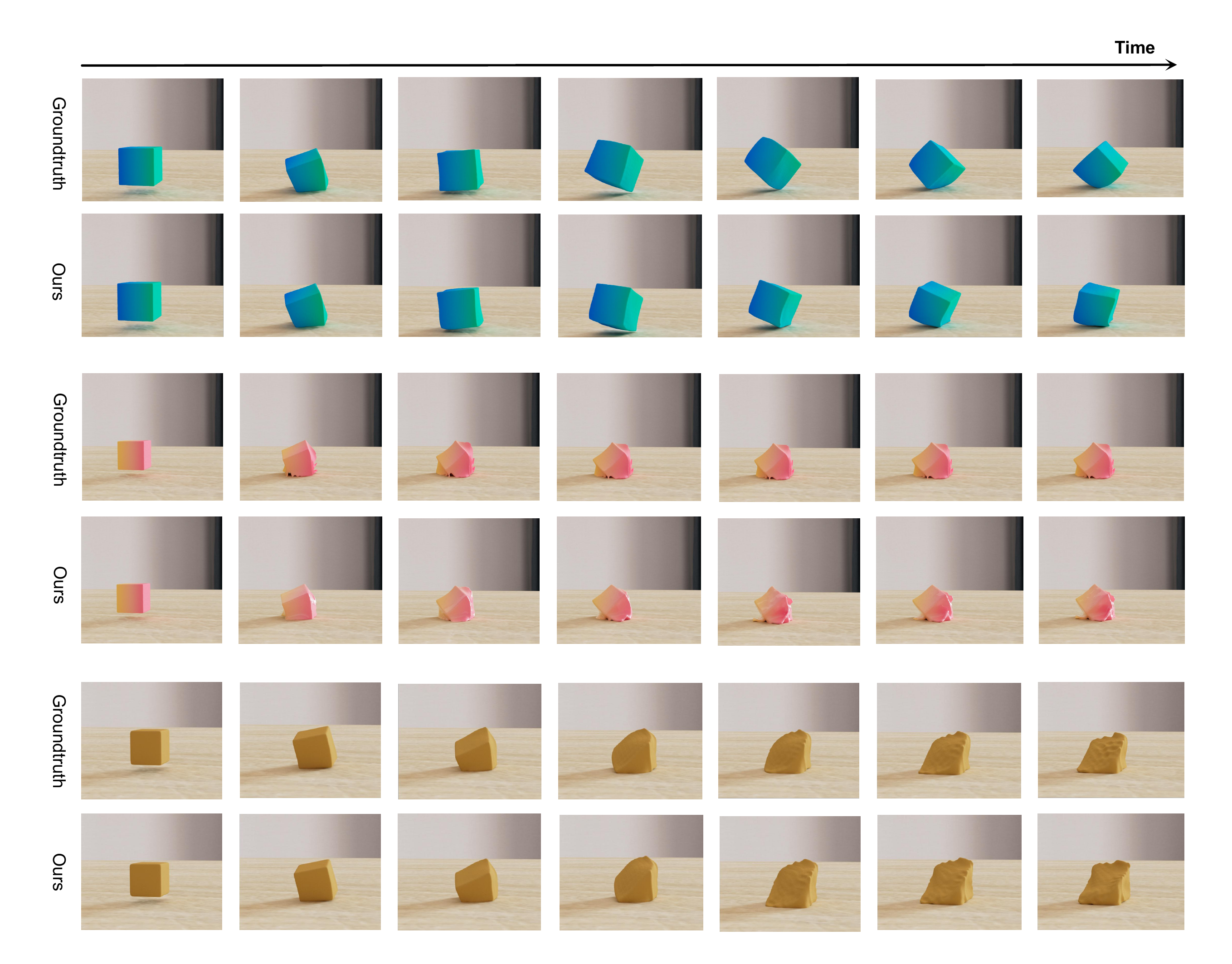}    \caption{\textbf{Qualitative Results} (via \textbf{PC-NCLaws-M} training strategy). We present visual results for the double-time generalization.}
    \label{geo}
\end{figure*}

\end{document}